\documentclass[12pt]{article}
\usepackage[table]{xcolor}
\usepackage{colortbl}
\definecolor{lightgray}{gray}{0.9}
\usepackage{graphicx}
\usepackage{lscape}
\usepackage{citesort}
\usepackage{amssymb}
\usepackage{appendix}
\usepackage{multirow}

\usepackage{graphicx}
\usepackage{lscape}
\usepackage{citesort}
\usepackage{amssymb}
\usepackage{appendix}
\usepackage{multirow}

\usepackage{mathrsfs}
\begin{document}
\def\qq{\langle \bar q q \rangle}
\def\uu{\langle \bar u u \rangle}
\def\dd{\langle \bar d d \rangle}
\def\sp{\langle \bar s s \rangle}
\def\GG{\langle g_s^2 G^2 \rangle}
\def\Tr{\mbox{Tr}}
\def\figt#1#2#3{
        \begin{figure}
        $\left. \right.$
        \vspace*{-2cm}
        \begin{center}
        \includegraphics[width=10cm]{#1}
        \end{center}
        \vspace*{-0.2cm}
        \caption{#3}
        \label{#2}
        \end{figure}
    }

\def\figb#1#2#3{
        \begin{figure}
        $\left. \right.$
        \vspace*{-1cm}
        \begin{center}
        \includegraphics[width=10cm]{#1}
        \end{center}
        \vspace*{-0.2cm}
        \caption{#3}
        \label{#2}
        \end{figure}
                }

\def\ds{\displaystyle}
\def\beq{\begin{equation}}
\def\eeq{\end{equation}}
\def\bea{\begin{eqnarray}}
\def\eea{\end{eqnarray}}
\def\beeq{\begin{eqnarray}}
\def\eeeq{\end{eqnarray}}
\def\ve{\vert}
\def\vel{\left|}
\def\ver{\right|}
\def\nnb{\nonumber}
\def\ga{\left(}
\def\dr{\right)}
\def\aga{\left\{}
\def\adr{\right\}}
\def\lla{\left<}
\def\rra{\right>}
\def\rar{\rightarrow}
\def\lrar{\leftrightarrow}
\def\nnb{\nonumber}
\def\la{\langle}
\def\ra{\rangle}
\def\ba{\begin{array}}
\def\ea{\end{array}}
\def\tr{\mbox{Tr}}
\def\ssp{{\Sigma^{*+}}}
\def\sso{{\Sigma^{*0}}}
\def\ssm{{\Sigma^{*-}}}
\def\xis0{{\Xi^{*0}}}
\def\xism{{\Xi^{*-}}}
\def\qs{\la \bar s s \ra}
\def\qu{\la \bar u u \ra}
\def\qd{\la \bar d d \ra}
\def\qq{\la \bar q q \ra}
\def\gGgG{\la g^2 G^2 \ra}
\def\q{\gamma_5 \not\!q}
\def\x{\gamma_5 \not\!x}
\def\g5{\gamma_5}
\def\sb{S_Q^{cf}}
\def\sd{S_d^{be}}
\def\su{S_u^{ad}}
\def\sbp{{S}_Q^{'cf}}
\def\sdp{{S}_d^{'be}}
\def\sup{{S}_u^{'ad}}
\def\ssp{{S}_s^{'??}}

\def\sig{\sigma_{\mu \nu} \gamma_5 p^\mu q^\nu}
\def\fo{f_0(\frac{s_0}{M^2})}
\def\ffi{f_1(\frac{s_0}{M^2})}
\def\fii{f_2(\frac{s_0}{M^2})}
\def\O{{\cal O}}
\def\sl{{\Sigma^0 \Lambda}}
\def\es{\!\!\! &=& \!\!\!}
\def\ap{\!\!\! &\approx& \!\!\!}
\def\md{\!\!\!\! &\mid& \!\!\!\!}
\def\ar{&+& \!\!\!}
\def\ek{&-& \!\!\!}
\def\kek{\!\!\!&-& \!\!\!}
\def\cp{&\times& \!\!\!}
\def\se{\!\!\! &\simeq& \!\!\!}
\def\eqv{&\equiv& \!\!\!}
\def\kpm{&\pm& \!\!\!}
\def\kmp{&\mp& \!\!\!}
\def\mcdot{\!\cdot\!}
\def\erar{&\rightarrow&}
\def\olra{\stackrel{\leftrightarrow}}
\def\ola{\stackrel{\leftarrow}}
\def\ora{\stackrel{\rightarrow}}

\def\simlt{\stackrel{<}{{}_\sim}}
\def\simgt{\stackrel{>}{{}_\sim}}


\title{
         {\Large
                 {\bf
                     Strong $\Sigma_bNB$ and $\Sigma_cND$ coupling constants in QCD
                 }
         }
      }

\author{\vspace{1cm}\\
{\small K. Azizi$^a$ \thanks {e-mail: kazizi@dogus.edu.tr}\,, Y.
Sarac$^b$
\thanks {e-mail: yasemin.sarac@atilim.edu.tr}\,\,, H.
Sundu$^c$ \thanks {e-mail: hayriye.sundu@kocaeli.edu.tr}} \\
{\small $^a$  Department of Physics, Do\u gu\c s University, Ac{\i}badem-Kad{\i}k\"oy, 34722 Istanbul, Turkey} \\
{\small $^b$ Electrical and Electronics Engineering Department,
Atilim University, 06836 Ankara, Turkey} \\
{\small $^c$ Department of Physics, Kocaeli University, 41380 Izmit,
Turkey}}
\date{}

\begin{titlepage}
\maketitle
\thispagestyle{empty}

\begin{abstract}
We study  the strong interactions   among the heavy bottom spin--1/2
$\Sigma_b$ baryon, nucleon and $B$ meson as well as the heavy charmed spin--1/2  $\Sigma_c$ baryon, nucleon and
$D$ meson in the context of QCD sum rules. We calculate  the corresponding
strong coupling form factors defining these vertices  by using a three point correlation function. We obtain the numerical values of the corresponding
strong coupling constants via the most prominent  structure entering the calculations.

\end{abstract}

~~~PACS number(s): 13.30.-a,  13.30.Eg, 11.55.Hx
\end{titlepage}

\section{Introduction}

In the recent years, substantial experimental improvements have been
made on the spectroscopic and decay properties of heavy hadrons,
which were accompanied by theoretical studies on various properties
of these hadrons. The mass spectrum of the baryons containing heavy
quark has been studied using different methods. The necessity of a
deeper understanding of heavy flavor physics requires a
comprehensive study on the processes of these baryons such as their
radiative, strong and weak decays. For some of related studies one
can refer to references \cite{Wang a2,Ebert,Faessler,Navarra,Azizi,Wang,Khodjamirian,Nieves,sarac,Gutsche}.

The investigation of the strong decays of heavy baryons can help us get valuable information on the perturbative and non-perturbative natures of QCD.
 The strong coupling constants defining such decays play important role in describing
the strong interaction among the heavy baryons and other
participated particles. Therefore, accurate determination of these
coupling constants enhance our understanding on the interactions as
well as the nature and structure of the participated particles. The
present work is an extension of our previous study on the coupling
constants $g_{\Lambda_bNB}$ and $g_{\Lambda_cND}$ \cite{Azizi14}.
Here, we study the strong interactions   among the heavy bottom
spin--1/2 $\Sigma_b$ baryon, nucleon and $B$ meson as well as the
heavy charmed spin--1/2  $\Sigma_c$ baryon, nucleon and $D$ meson in
the context of QCD sum rules. In particular, we
calculate the strong coupling constants $g_{\Sigma_bNB}$ and
$g_{\Sigma_cND}$. These  coupling constants together with the
$g_{\Lambda_bNB}$
 and $g_{\Lambda_cND}$ discussed in our previous work, may also be used   in the
bottom and charmed mesons clouds description of the nucleon which
can be used to explain the exotic events observed by different
Collaborations. In addition, the determination of the properties of
the $B$ and $D$ mesons in nuclear medium requires the consideration
of their interactions  with the nucleons from which the
$\Lambda_{b[c]}$ and $\Sigma_{b[c]}$ are produced. Therefore, to
determine the modifications on the masses, decay constants and other
parameters of the $B$ and $D$ mesons in nuclear medium, one needs to
consider the contributions of the baryons $\Sigma_{b[c]}$ together
with the $\Lambda_{b[c]}$  and have the values of the strong
coupling constants $g_{\Sigma_bNB}$ and $g_{\Sigma_cND}$ besides the
couplings $g_{\Lambda_bNB}$ and $g_{\Lambda_cND}$
~\cite{Azizi1,Kumar,Wang1,Hayashigaki}. In the  literature,
one can unfortunately find only a few works on the strong couplings
of the heavy baryons with the nucleon and heavy mesons. One
approximate prediction for the strong coupling $g_{\Lambda_cND}$ was
made at zero transferred momentum squared \cite{Navarra}. The strong
couplings of the charmed baryons with the nucleon and $D$ meson were
also discussed in \cite{Khodjamirian} in the framework of light cone
QCD sum rules.

This paper is organized in three sections as follows.  In the next section, we present the  details
of the calculations of the strong coupling  form factors among the particles under
consideration. In section 3, the numerical analysis of the obtained
sum rules and discussions about the results are presented.

\section{ Theoretical framework}
This section is devoted to  the details of the calculations of the
strong coupling form factors $g_{\Sigma_b NB}(q^2)$ and
$g_{\Sigma_cND}(q^2)$ from which  the strong coupling constants
among the participating particles are obtained at
$Q^2=-q^2=-m_{B[D]}^2$, subsequently. In order to accomplish this
purpose, the following three-point correlation function is used:
\begin{eqnarray}\label{CorrelationFunc1}
\Pi=i^2 \int d^4x~ \int d^4y~e^{-ip\cdot x}~ e^{ip^{\prime}\cdot
y}~{\langle}0| {\cal T}\left ( \eta_{N}(y)~\eta_{B[D]}(0)~
\bar{\eta}_{\Sigma_b[\Sigma_c]}(x)\right)|0{\rangle},
\end{eqnarray}
whith ${\cal T}$ being the time ordering operator and $q=p-p'$ is
the transferred momentum. The currents  $\eta_{N}$, $\eta_{B[D]}$ and $\eta_{\Sigma_b[\Sigma_c]}$ presented in
Eq.~(\ref{CorrelationFunc1}) correspond to the the interpolating
currents of the $N$, $B[D]$ and $\Sigma_{b[c]}$, respectively and
their explicit expressions can be given in terms of the quark field
operators as
\begin{eqnarray}\label{InterpolatingCurrents}
\eta_{\Sigma_{b}[\Sigma_{c}]}(x)&=&\varepsilon_{ijk}\Big(u^{i^T}(x)C\gamma_{\mu}d^{j}(x)\Big)\gamma_5
\gamma_{\mu}b[c]^{k}(x),
\nonumber \\
\eta_{N}(y)&=&\varepsilon_{ijk}\Big(u^{i^T}(y)C\gamma_{\mu}u^{j}(y)\Big)\gamma_{5}\gamma_{\mu}d^{k}(y),
\nonumber \\
\eta_{B[D]}(0)&=&\bar{u}(0)\gamma_5b[c](0),
\end{eqnarray}
where $C$ denotes the charge conjugation operator; and $i$, $j$ and
$k$ are color indices.

In the course of calculation of the three-point correlation function
one follows two different ways. The first way is called as OPE side
and the calculation is made in deep Euclidean region in terms of
quark and gluon degrees of freedom using the operator product
expansion. The second way is called as hadronic side and the
hadronic degrees of freedoms are considered to perform this side of
the calculation. The QCD sum rules for the coupling form factors are
attained via the match of these two sides. The contributions of the
higher states and continuum are suppressed by a double Borel
transformation applied to both sides with respect to the variables
$p^2$ and $p'^2$.

\subsection{OPE Side}

For the calculation of the OPE side of the  correlation function
which is done in deep Euclidean region, where $p^2\rightarrow
-\infty$ and $p'^2\rightarrow -\infty$, one puts the interpolating
currents given in Eq.~(\ref{InterpolatingCurrents}) into the
correlation function, Eq.~(\ref{CorrelationFunc1}). Possible
contractions of all quark pairs via Wick's theorem leads to
\begin{eqnarray}\label{correlfuncOPE1}
\Pi^{OPE}&=&i^2\int d^{4}x\int d^{4}ye^{-ip\cdot
x}e^{ip^{\prime}\cdot y}\varepsilon_{abc}\varepsilon_{ij\ell}
\nonumber \\
&\times&
\Bigg\{\gamma_5\gamma_{\nu}S^{cj}_{d}(y-x)\gamma_{\mu}CS_{u}^{bi^T}(y-x)C\gamma_{\nu}S^{ah}_{u}(y)
\gamma_{5}S_{b[c]}^{h\ell}(-x)\gamma_{\mu}\gamma_5
\nonumber \\
&-&\gamma_5\gamma_{\nu}S^{cj}_{d}(y-x)\gamma_{\mu}CS_{u}^{ai^T}(y-x)C\gamma_{\nu}S^{bh}_{u}(y)
\gamma_{5}S_{b[c]}^{h\ell}(-x)\gamma_{\mu}\gamma_5
 \Bigg\}~,
\end{eqnarray}
where $ S_{b[c]}(x)$ and $S_{u[d]}(x)$  are the heavy and light quark
propagators  whose explicit forms
can be found in Refs.~\cite{Reinders,Azizi14}.

After some straightforward calculations (for details refer to the
Ref.~\cite{Azizi14}), the correlation function in OPE side comes out
in terms of different Dirac structures as
\begin{eqnarray}\label{correlfuncOPE1Last}
\Pi^{OPE}&=&\Pi_1(q^2)\gamma_5+\Pi_2(q^2)\!\not\!{p}\gamma_5+\Pi_3(q^2)\!\not\!{q}\!\not\!{p}\gamma_5
+\Pi_4(q^2)\!\not\!{q}\gamma_5.
\end{eqnarray}
Each $\Pi_i(q^2)$ function involves the perturbative and
non-perturbative parts and is written as
\begin{eqnarray}\label{QCDside1}
\Pi_i(q^2)=\int^{}_{}ds\int^{}_{}ds^{\prime}
\frac{\rho_i^{pert}(s,s^{\prime},q^2)+\rho_i^{non-pert}(s,s^{\prime},q^2)}{(s-p^2)
(s^{\prime}-p^{\prime^2})}~.
\end{eqnarray}
The spectral densities, $\rho_i(s,s',q^2)$, appearing in
Eq.~(\ref{QCDside1}) are obtained from the imaginary parts of the
$\Pi_{i}$ functions, i.e.,
$\rho_i(s,s',q^2)=\frac{1}{\pi}Im[\Pi_{i}]$. Here to provide
examples of the explicit forms of the spectral densities, among the
Dirac structures presented above, we only present the results
obtained for the Dirac structure $\!\not\!{p}\gamma_5$, that is
$\rho_2^{pert}(s,s^{\prime},q^2)$ and
$\rho_2^{non-pert}(s,s^{\prime},q^2)$, which are obtained as
\begin{eqnarray}\label{rho1pert}
\rho_2^{pert}(s,s^{\prime},q^2)&=& \int_{0}^{1}dx \int_{0}^{1-x}dy
\frac{1}{32\pi^4(x+y-1)^2}\Bigg\{2m_{b[c]}^3x\Big(3x^2-y-2x+3xy\Big)
\nonumber \\
&-&3m_{b[c]}^2x(x+y-1)
\Big[-m_u(x+4y-2)+m_d(3x+6y-2)\Big]
\nonumber \\
&-&q^2m_{b[c]}x\Big[y-3y^2+x^2(8y-1)+x(1-6y+8y^2)\Big]+m_{b[c]}(x+y-1)
\nonumber \\
&\times&
\Big[sx(8x^2-3y-5x+8xy)+s^{\prime}\Big(8xy^2-5xy-2y^2-x^2+8x^2y\Big)\Big]
\nonumber \\
&-&(x+y-1)\Big[3m_ds\Big(4x^3+y-y^2-7x^2+12x^2y+3x-12xy+8xy^2\Big)
\nonumber \\
&+&3m_dq^2y\Big(4x-4x^2+y-8xy\Big)
+3m_ds^{\prime}y\Big(3-7x+4x^2-11y+12xy+8y^2\Big)\Big]
\nonumber \\
&-&m_us\Big[4x^3+y(3-2y)+x^2(20y-13)
+x(9-27y+16y^2)\Big]
\nonumber \\
&+&m_us^{\prime}y\Big(9-14x+4x^2-26y+20xy+16y^2\Big)
\nonumber \\
&+& m_uq^2y(11x-4x^2+2y-16xy-3)
\Bigg\}\Theta\Big[L_2(s,s^{\prime},q^2)\Big] ,
\end{eqnarray}
and
\begin{eqnarray}\label{rho1nonpert}
\rho_2^{non-pert}(s,s^{\prime},q^2)&=&\Big\{\frac{\langle u
\bar{u}\rangle}{24\pi^2(q^2-m_{b[c]}^2)}\Big(3m_{b[c]}m_d-3m_um_d+3m_u^2-q^2+s-s^{\prime}\Big)
\nonumber \\
&+&\langle
\alpha_s\frac{G^2}{\pi}\rangle\frac{m_{b[c]}\Big(3q^2-3m_{b[c]}^2-2s^{\prime}\Big)}{192\pi^2(q^2-m_{b[c]}^2)^2}
+\frac{m_0^2\langle u \bar{u}\rangle}{16\pi^2(q^2-m_b^2)}
\Bigg\}\Theta\Big[L_1(s,s^{\prime},q^2)\Big]
\nonumber \\
&-&\Big(\langle d \bar{d}\rangle-\langle u
\bar{u}\rangle\Big)\int_{0}^{1}dx \int_{0}^{1-x}dy\frac{3x+6y-2}{4\pi^2}\Theta\Big[L_2(s,s^{\prime},q^2)\Big],
\end{eqnarray}
where $\Theta[...]$ stands for the unit-step function and  $L_1(s,s^{\prime},q^2)$ and $L_2(s,s^{\prime},q^2)$
are defined as
\begin{eqnarray}\label{teta1}
L_1(s,s^{\prime},q^2)&=&s^{\prime},
\nonumber \\
L_2(s,s^{\prime},q^2)&=&-m_{b[c]}^2x+sx-sx^2+s^{\prime}y+q^2xy-sxy-s^{\prime}xy-s^{\prime}y^2.
\end{eqnarray}
%
%

\subsection{Hadronic Side}

 On the hadronic side, considering the quantum numbers of the interpolating fields one place the complete sets of
 intermediate $\Sigma_{b}[\Sigma_{c}]$, $B[D]$ and $N$ hadronic states into the correlation function. After carrying out the four-integrals, we get
\begin{eqnarray} \label{physide1}
\Pi^{HAD}&=&\frac{\langle 0 \mid
 \eta_{N}\mid N(p^{\prime})\rangle \langle 0 \mid
 \eta_{B[D]}\mid B[D](q)\rangle \langle
\Sigma_{b}[\Sigma_{c}](p) \mid
 \bar{\eta}_{\Sigma_{b}[\Sigma_{c}]}\mid 0\rangle }{(p^2-m_{\Sigma_{b}[\Sigma_{c}]}^2)(p^{\prime^2}-m_{N}^2)(q^2-m_{B[D]}^2)}
 \nonumber \\
&\times&\langle N(p^{\prime})B[D](q)\mid
\Sigma_{b}[\Sigma_{c}](p)\rangle+\cdots~.
\end{eqnarray}
In the above equation, the contributions of the higher states and
continuum are denoted by $\cdots$ and the matrix elements are
represented in terms of the hadronic parameters as follows:
\begin{eqnarray}\label{matriselement1}
\langle 0 \mid
 \eta_{N}\mid N(p^{\prime})\rangle&=&\lambda_N u_N(p^{\prime}, s^{\prime}),
\nonumber \\
\langle \Sigma_b(p) \mid
 \bar{\eta}_{\Sigma_b[\Sigma_c]}\mid
 0\rangle&=&\lambda_{\Sigma_b[\Sigma_c]}\bar{u}_{\Sigma_b[\Sigma_c]}(p, s),
\nonumber \\
\langle 0 \mid
 \eta_{B[D]}\mid B[D](q)\rangle&=&i\frac{m_{B[D]}^2f_{B[D]}}{m_u+m_{b[c]}},
\nonumber \\
\langle N(p^{\prime})B[D](q)\mid
 \Sigma_b[\Sigma_c](p)\rangle&=&g_{\Sigma_bNB[\Sigma_c ND]}\bar{u}_N(p^{\prime},s^{\prime})i
 \gamma_5 u_{\Sigma_b[\Sigma_c]}(p, s).
\end{eqnarray}
Here  $\lambda_N$ and $\lambda_{\Sigma_b[\Sigma_c]}$ are
 residues of the $N$ and $\Sigma_b[\Sigma_c]$ baryons, respectively,  $f_{B[D]}$ is the leptonic decay constant of
$B[D]$ meson and $g_{\Sigma_bNB[\Sigma_cND]}$ is the strong coupling
form factor among $\Sigma_b[\Sigma_c]$, $N$ and $B[D]$ particles.
%
Using  Eq.~(\ref{matriselement1}) in  Eq.~(\ref{physide1}) and summing over the spins of the particles,
we obtain
\begin{eqnarray} \label{physide1Last}
\Pi^{HAD}&=&i^2\frac{m_{B[D]}^2f_{B[D]}}{m_{b[c]}+m_u}\frac{\lambda_N
\lambda_{\Sigma_b [\Sigma_c]} g_{\Sigma_bNB[\Sigma_c
ND]}}{(p^2-m_{\Sigma_b[\Sigma_c]}^2)(p^{\prime^2}-m_N^2)(q^2-m_{B[D]}^2)}
\nonumber \\
&\times&\Big\{(m_Nm_{\Sigma_b[\Sigma_c]}-
m_{\Sigma_b[\Sigma_c]}^2)\gamma_5+(m_{\Sigma_b[\Sigma_c]}-m_N)\not\!p\gamma_5+
\not\!q\not\!p\gamma_5-m_{\Sigma_b[\Sigma_c]}
\not\!q\gamma_5\Big\}\nonumber\\
&+&\cdots~.
\end{eqnarray}
To acquire the final form of the hadronic side of the correlation
function we perform the double Borel transformation with respect to
the initial and final momenta squared,
\begin{eqnarray} \label{physide1Last1}
\widehat{\textbf{B}}\Pi^{HAD}&=&i^2\frac{m_{B[D]}^2f_{B[D]}}{m_{b[c]}+m_u}\frac{\lambda_N
\lambda_{\Sigma_b[\Sigma_c]}
g_{\Sigma_bNB[\Sigma_cND]}}{(q^2-m_{B[D]}^2)}e^{-\frac{m_{\Sigma_b[\Sigma_c]}^2}{M^2}}
e^{-\frac{m_N^2}{M^{\prime^2}}}
\nonumber \\
&\times&\Big\{(m_Nm_{\Sigma_b[\Sigma_c]}-
m_{\Sigma_b[\Sigma_c]}^2)\gamma_5+(m_{\Sigma_b[\Sigma_c]}-m_N)\not\!p\gamma_5+\not\!q\not\!p\gamma_5
-m_{\Sigma_b[\Sigma_c]}
\not\!q\gamma_5\Big\}\nonumber\\&+&\cdots~,
\end{eqnarray}
where $M^2$ and $M^{\prime^2}$ are Borel mass parameters.

As it was already  stated, the match of the hadronic and OPE sides
of the correlation function in Borel scheme provides us with the  QCD sum rules for
the strong form factors. The consequence of that match for
$\!\not\!{p}\gamma_5$ structure leads us to
\begin{eqnarray}\label{couplingconstant}
&&g_{\Sigma_{b}NB[\Sigma_{c}ND]}(q^2)=-e^{\frac{m_{\Sigma_b[\Sigma_c]}^2}{M^2}}e^{
\frac{m_N^2}{M^{\prime^2}}}~
\frac{(m_{b[c]}+m_u)(q^2-m_{B[D]}^2)}{m_{B[D]}^2f_{B[D]}\lambda_{\Sigma_{b}[\Sigma_c]}^{\dag}
\lambda_N(m_Nm_{\Sigma_{b}[\Sigma_c]}- m_{\Sigma_b[\Sigma_c]}^2)}
\nonumber \\
&\times&
\Bigg\{\int^{s_0}_{(m_{b[c]}+m_{u}+m_{d})^2}ds\int^{s'_0}_{(2m_u+m_d)^2}ds^{\prime}e^{-\frac{s}{M^2}}
e^{-\frac{s^{\prime}}{M^{\prime^2}}}
\Big[\rho_2^{pert}(s,s^{\prime},q^2)+\rho_2^{non-pert}(s,s^{\prime},q^2)\Big]\Bigg\}~,\nonumber\\
\end{eqnarray}
where $s_0$ and $s'_0$ are continuum thresholds in
$\Sigma_b[\Sigma_c]$ and $N$ channels, respectively.

\section{Numerical analysis}

Having obtained the QCD sum rules for the strong coupling form
factors, in this section, we present the numerical analysis of our
results and discuss the dependence of the strong coupling form
factors under consideration on $Q^2=-q^2$. To this aim, beside the
input parameters given in table 1, one needs to determine the
working intervals of four auxiliary parameters $M^2$, $M'^2$, $s_0$
and $s'_0$. These parameters originate from the double Borel
transformation and continuum subtraction. The determination of the
working regions of them is made on the basis of that the results
obtained for the strong coupling form factors  be roughly
independent of these helping parameters.

The continuum thresholds  $s_{0}$ and $s'_0$ are the parameters
related to the beginning of the continuum in the initial and final
channels. If the ground state masses are given by $m$ and $m'$ for
the initial and final channels, respectively, to excite the particle to
the first excited state having the same quantum numbers with them
one needs to provide the energies $\sqrt{s_0}-m$ and
$\sqrt{s'_0}-m'$. For the considered transitions, these quantities
can be determined from well known excited states of the initial and
final states \cite{Olive} which are roughly in between
$0.1~\mbox{GeV}$ and $0.3~\mbox{GeV}$. From these intervals, the
working regions for the continuum thresholds are determined as
$34.9[6.5]~\mbox{GeV$^2$}\leq s_0\leq37.4[7.6]~\mbox{GeV$^2$}$ and
$1.04~\mbox{GeV$^2$}\leq s'_0\leq1.99~\mbox{GeV$^2$}$ for the vertex
$\Sigma_bNB[\Sigma_cND]$.

\begin{table}[ht]\label{Table1}
\centering \rowcolors{1}{lightgray}{white}
\begin{tabular}{cc}
\hline \hline
   Parameters  &  Values
           \\
\hline \hline
$m_{b}$              & $(4.18\pm0.03)~\mbox{GeV}$\cite{Olive}\\
$m_{c}$              & $(1.275\pm0.025)~\mbox{GeV}$\cite{Olive}\\
$m_{d}$              & $4.8^{+0.5}_{-0.3}~\mbox{MeV}$\cite{Olive}\\
$ m_{u} $            &$2.3^{+0.7}_{-0.5}~\mbox{MeV}$ \cite{Olive}\\
$ m_{B}$    &   $ (5279.26\pm0.17)~\mbox{MeV}$ \cite{Olive}  \\
$ m_{D}$    &   $ (1864.84\pm0.07)~\mbox{MeV}$ \cite{Olive}  \\
$ m_{N} $      &   $ (938.272046\pm0.000021)~\mbox{MeV}$  \cite{Olive} \\
$ m_{\Sigma_b} $      &   $ (5811.3\pm1.9) ~\mbox{MeV} $ \cite{Olive}  \\
$ m_{\Sigma_c} $      &   $(2452.9\pm 0.4) ~\mbox{MeV} $ \cite{Olive}  \\
$ f_{B} $      &   $(248\pm23_{exp}\pm25_{Vub}) ~\mbox{MeV}$
\cite{Khodjamirian1} \\
$ f_{D} $      &   $(205.8\pm8.5\pm2.5) ~\mbox{MeV}$  \cite{CLEO} \\
$ \lambda_{N}^2 $      &   $0.0011\pm0.0005  ~\mbox{GeV$^6$}$  \cite{Azizi2} \\
$ \lambda_{\Sigma_b} $      &   $(0.062\pm0.018)$ $\mbox{GeV$^3$}$  \cite{Azizi} \\
$ \lambda_{\Sigma_c} $      &   $(0.045\pm0.015)$ $\mbox{GeV$^3$}$  \cite{Azizi} \\
$  \langle \bar{u}u\rangle(1~GeV)=\langle \bar{d}d\rangle(1~GeV)$&
$-(0.24\pm0.01)^3 $ $\mbox{GeV$^3$}$
 \cite{Ioffe} \\
$ \langle\frac{\alpha_sG^2}{\pi}\rangle $       &   $(0.012\pm0.004)$ $~\mbox{GeV$^4$}$
\cite{belyaev}   \\
$ m_0^2(1~GeV) $       & $(0.8\pm0.2)$ $~\mbox{GeV$^2$}$
\cite{belyaev}   \\
 \hline \hline
\end{tabular}
\caption{Input parameters used in  calculations.}
\end{table}

Here, we shall comment on the selection of the most prominent Dirac structure to determine the corresponding strong coupling form factors. In principle, one can choose
any structure for determination of these strong coupling form factors. However, we should choose the most reliable one considering the following criteria:
\begin{itemize}
\item the pole/continuum should be the largest,
\item the series of sum rule should demonstrate the best convergence, i.e. the perturbative  part should have the largest contribution and the operator with highest dimension should have relatively small contribution.
\end{itemize}
Our numerical calculations show that these conditions lead to choose the structure  $\!\not\!{q}\!\not\!{p}\gamma_5$ as the most prominent structure. In the following, we will use this structure to numerically analyze the obtained sum rules.

Now, we proceed to present the Borel windows considering the selected structure. The working windows for the  Borel parameters $M^2$ and $M'^2$ are determined considering again the  pole dominance and
convergence of the OPE. By  requirement that the pole contribution
exceeds the contributions of the higher states and continuum, and
that the contribution of the perturbative part exceeds the
non-perturbative contributions we find the
windows $10[2]~\mbox{GeV$^2$}\leq M^2\leq 20[6]~\mbox{GeV$^2$}$ and
$1~\mbox{GeV$^2$}\leq M'^2\leq 3~\mbox{GeV$^2$}$ for the Borel mass
parameters of the strong vertex $\Sigma_bNB[\Sigma_cND]$.  For
these intervals, our results show weak dependence on the Borel mass
parameters (see figures 1-2).

\begin{figure}[h!]
\includegraphics[totalheight=6cm,width=8cm]{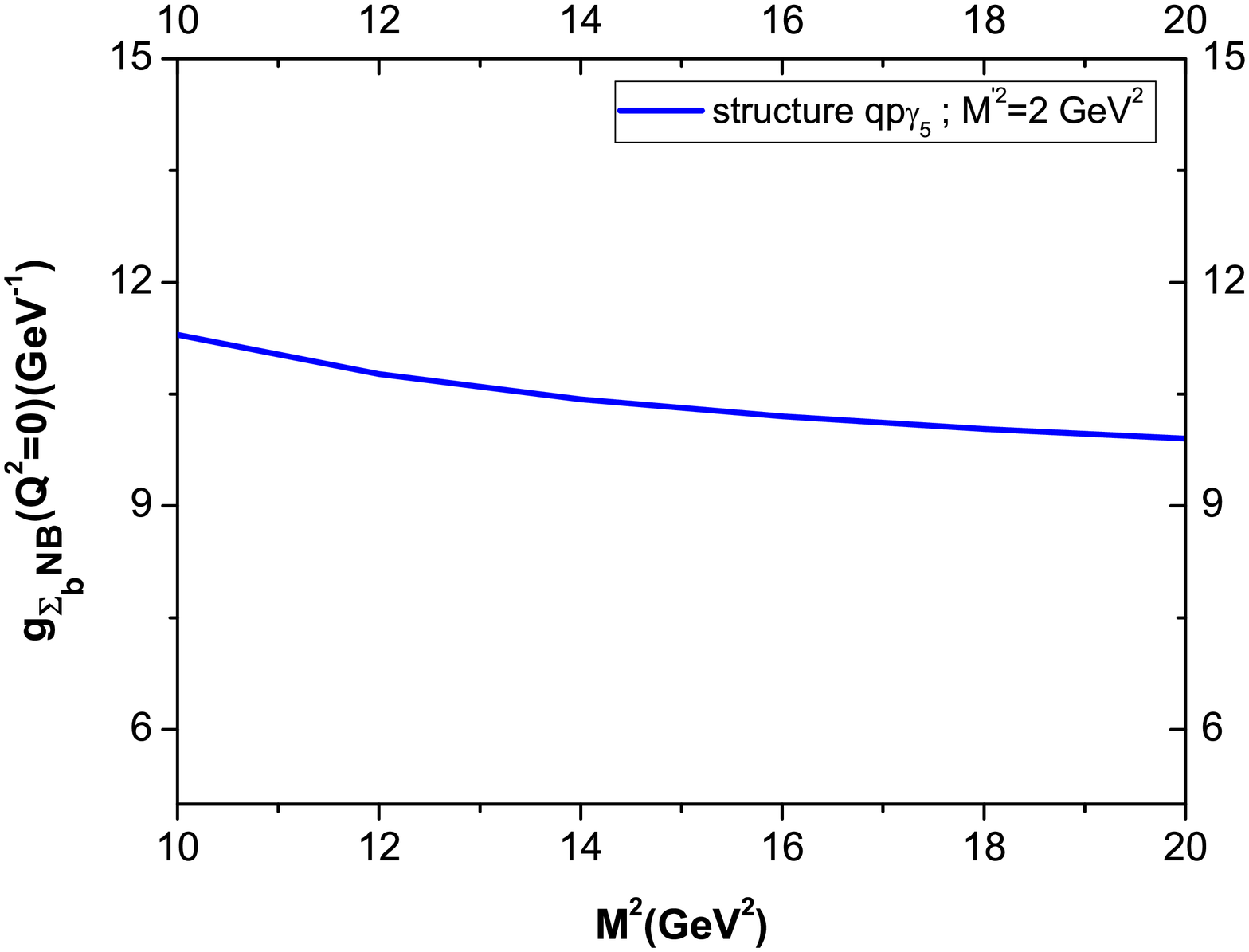}
\includegraphics[totalheight=6cm,width=8cm]{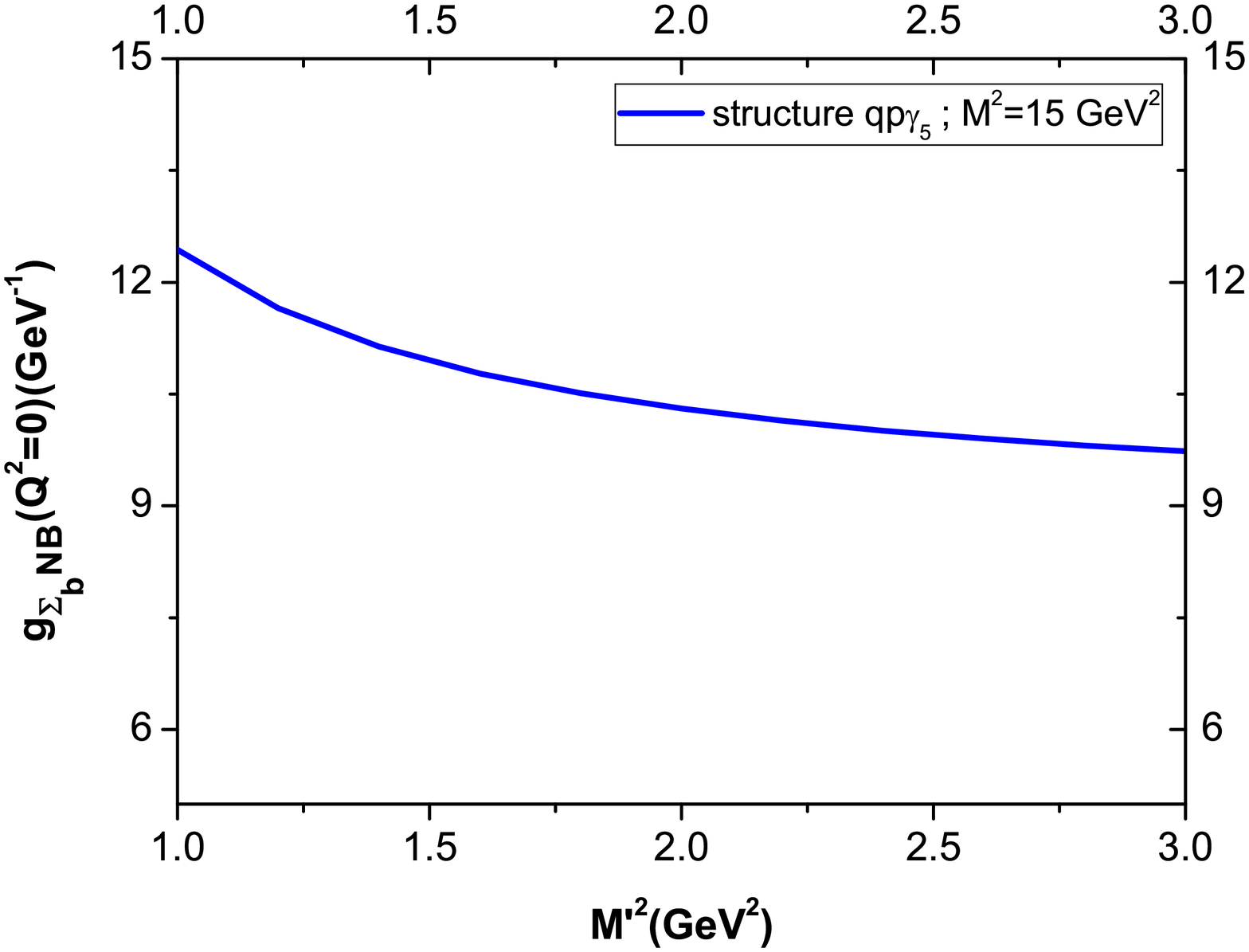}
\caption{\textbf{Left:} $g_{\Sigma_bNB}(Q^2=0)$ as a function of
the Borel mass $M^2$ at average values of continuum thresholds.
\textbf{Right:}
 $g_{\Sigma_bNB}(Q^2=0)$ as a function of the
Borel mass $M^{\prime^2}$ at average values of continuum
thresholds. } \label{gLamdabNBMsqMpsq}
\end{figure}
\begin{figure}[h!]
\includegraphics[totalheight=6cm,width=8cm]{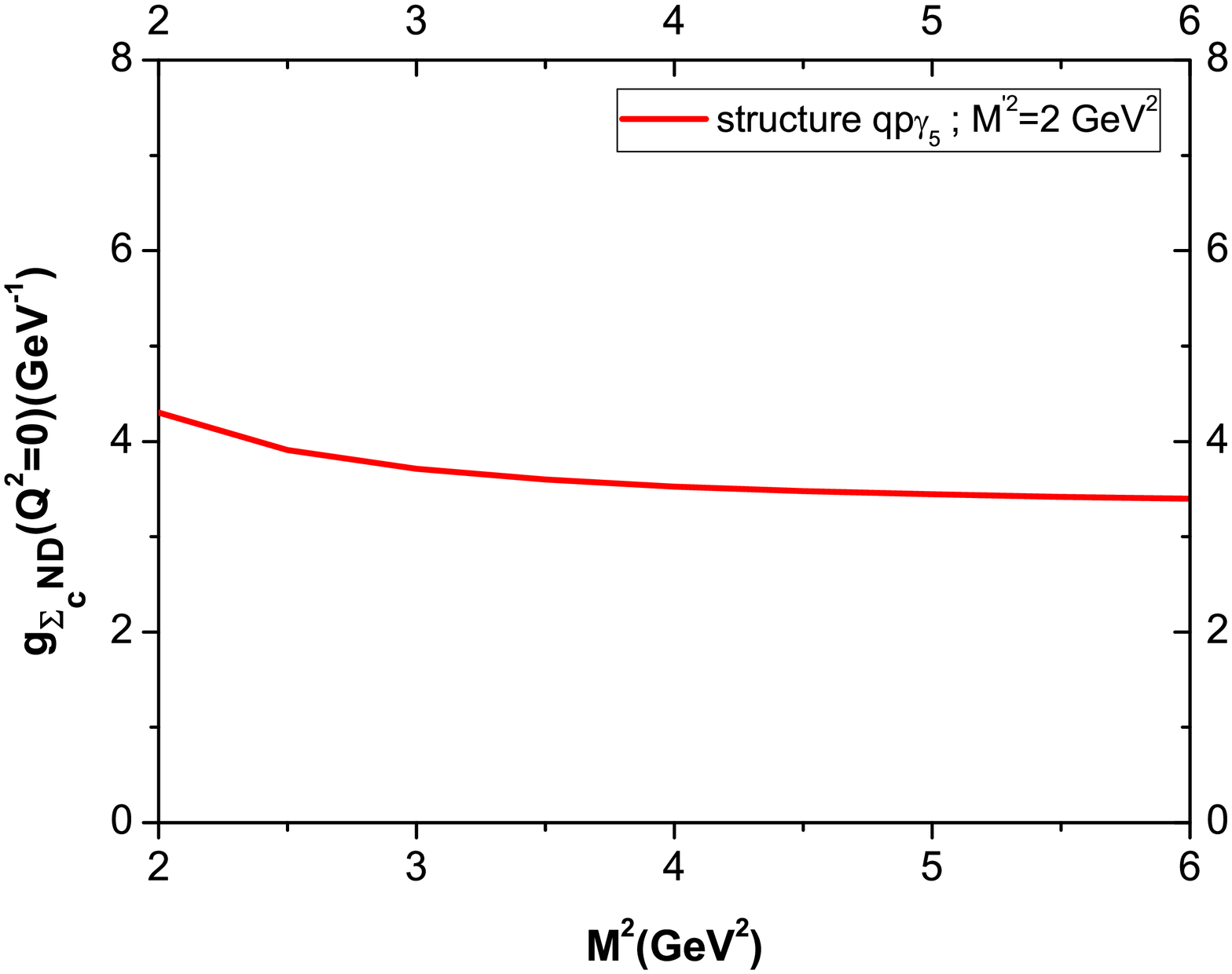}
\includegraphics[totalheight=6cm,width=8cm]{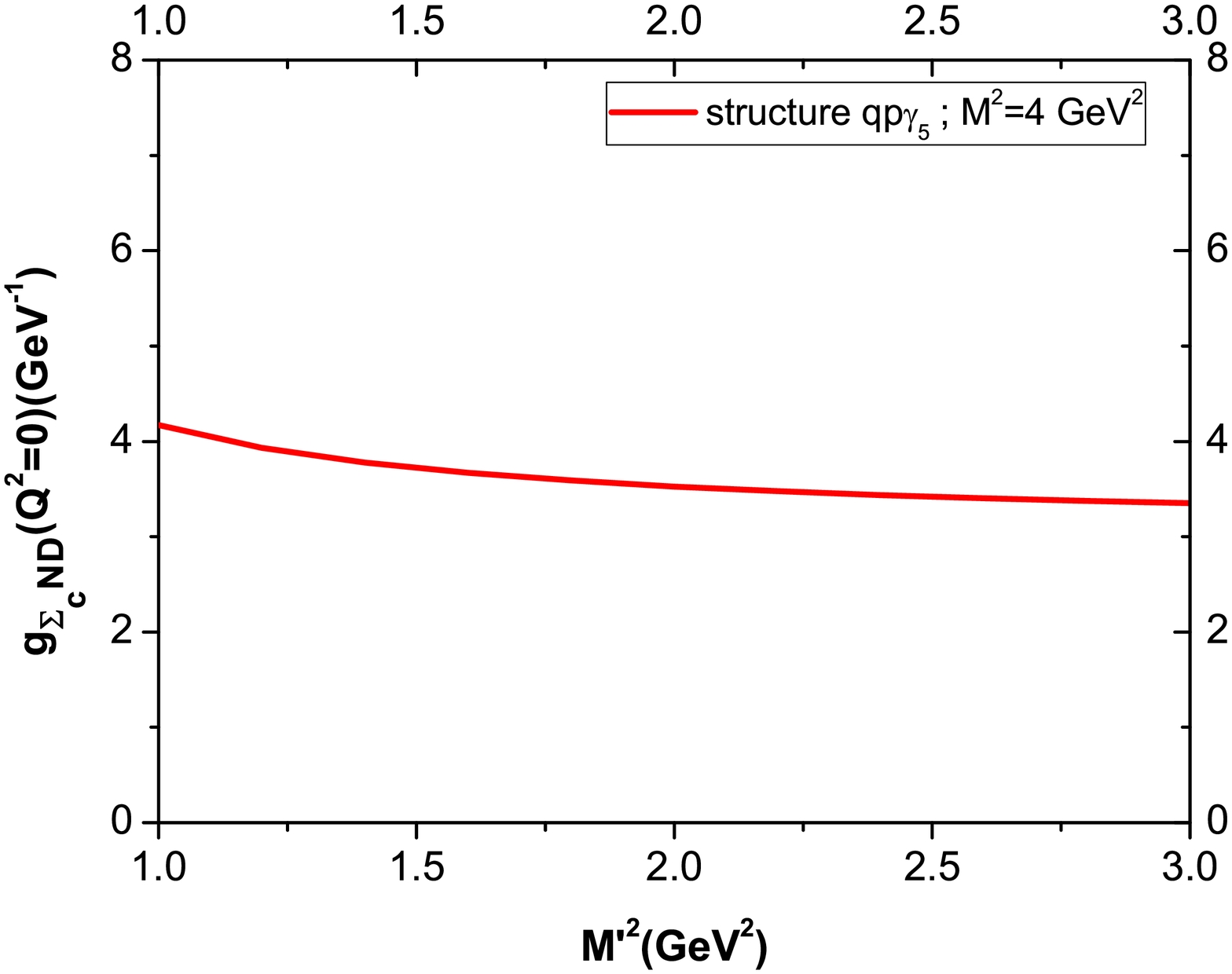}
\caption{The same as figure 1 but for $g_{\Sigma_cND}(Q^2=0)$. }
\label{gLamdabNBMsqMpsq}
\end{figure}

\begin{figure}[h!]
\includegraphics[totalheight=6cm,width=8cm]{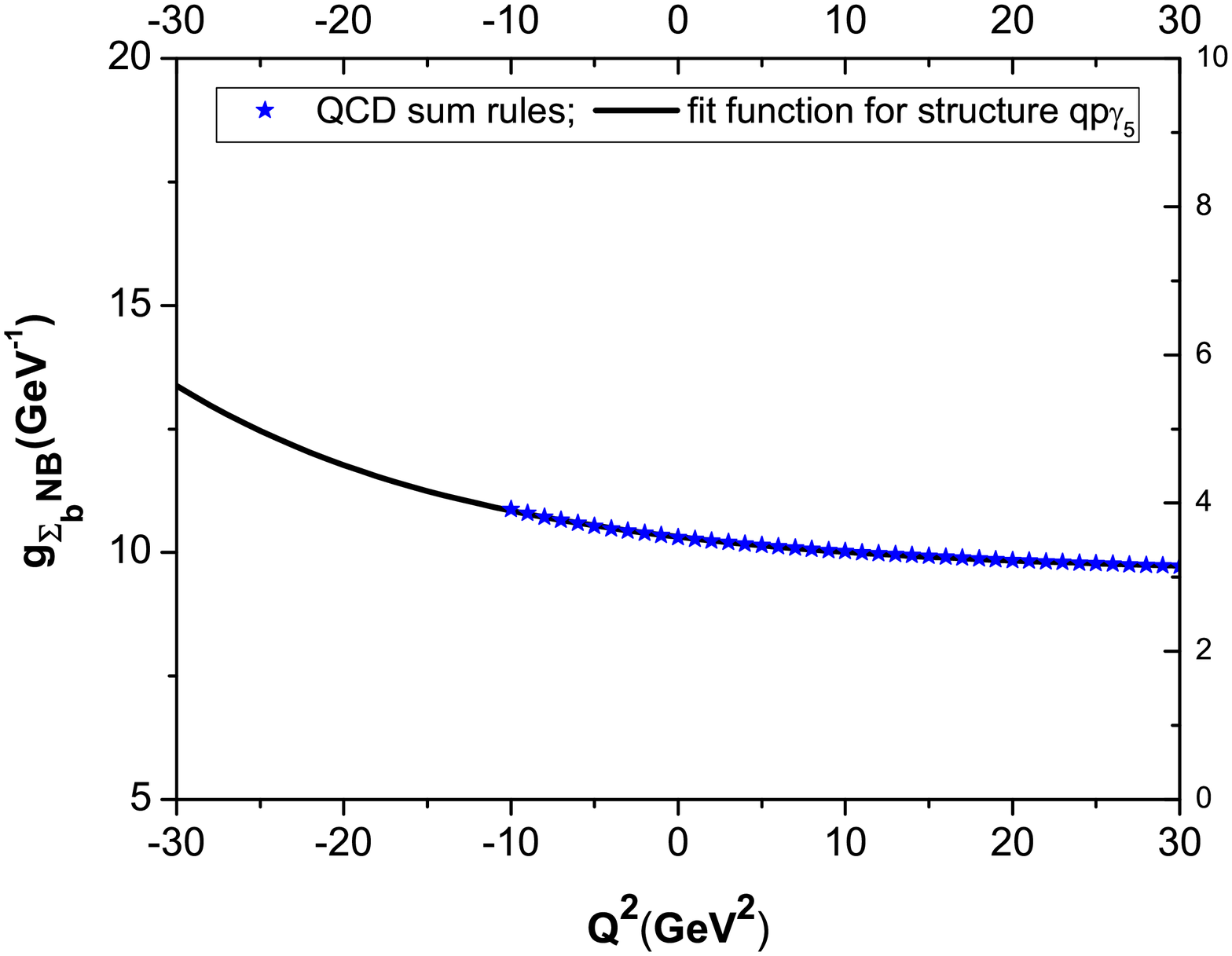}
\includegraphics[totalheight=6cm,width=8cm]{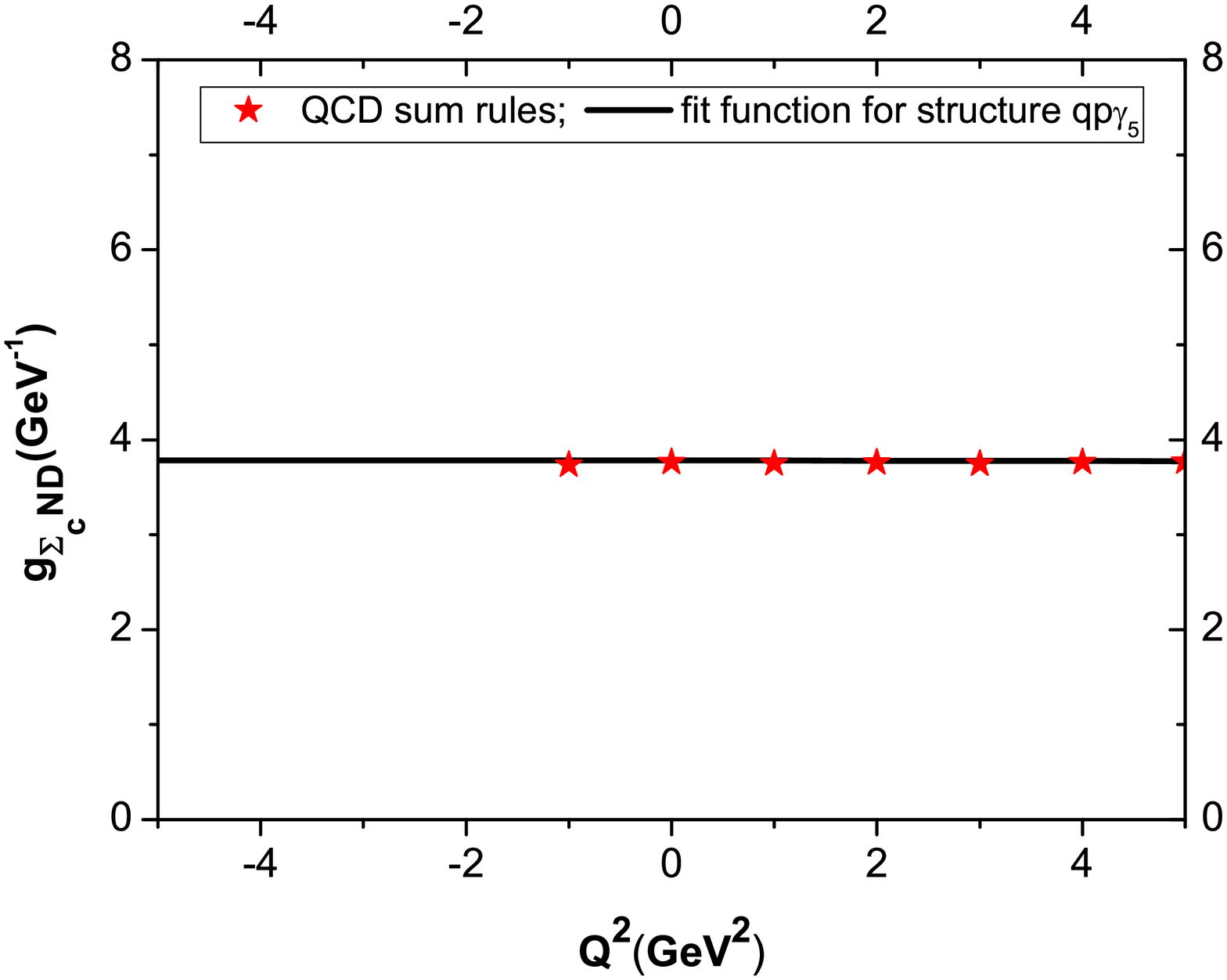}
\caption{\textbf{Left:} $g_{\Sigma_bNB}(Q^2)$ as a function of
 $Q^2$ at average values of the continuum thresholds and Borel mass parameters. \textbf{Right:}
 $g_{\Sigma_cND}(Q^2)$ as a function of  $Q^2$ at average values of the continuum thresholds and
 Borel mass parameters. } \label{gLamdabNBMsqMpsq}
\end{figure}

Subsequent to the determination of the auxiliary parameters, their
working windows together with the other input parameters are used to
ascertain the dependency of the strong coupling form factors on
$Q^2$. From our analysis we observe that the dependency of the
strong coupling form factors on $Q^2$ is well characterized by the
following fit function:
\begin{eqnarray}\label{fitfunc}
g_{\Sigma_bNB[\Sigma_cND]}(Q^2)=c_1\exp\Big[-\frac{Q^2}{c_2}\Big]+c_3.
\end{eqnarray}
The values of the parameters $c_1$, $c_2$ and $c_3$ 
 for $\Sigma_bNB$ and $\Sigma_cND$ can be seen
in  tables \ref{fitparam} and \ref{fitparam1}, respectively.
Considering the average values of the continuum thresholds and Borel
mass parameters we demonstrate the variation of the strong coupling
form factors with respect to  $Q^2$ for the QCD sum rules as well as the
fitting results in figure 3. The figure indicates the truncation of
the  QCD sum rules at some points at negative values of $Q^2$. It
can be seen from the figure that there is a good consistency among
the results obtained from the QCD sum rules and fit function up to
these points. The fit function is used to determine the value of the
strong coupling constant at $Q^2=-m_{B[D]}^2$,
and the results are presented in table~\ref{couplingconstant}. The
presented errors in these results originate from the uncertainties of
the input parameters together with the uncertainties coming from the
determination of the working regions of the auxiliary parameters.
%
\begin{table}[h]
\renewcommand{\arraystretch}{1.5}
\addtolength{\arraycolsep}{3pt}
$$
\begin{array}{|c|c|c|c|}
\hline \hline
       \mbox{Parameters}    & c_1 (\mbox{GeV$^{-1}$}) & c_2 (\mbox{GeV$^2$})&c_3 (\mbox{GeV$^{-1}$})    \\
\hline
  \mbox{Values} &0.72\pm 0.20&18.03\pm4.87&9.60\pm2.88 \\
                        \hline \hline
\end{array}
$$
\caption{Parameters appearing in the fit function of the coupling
form factor for $\Sigma_bNB$ vertex.} \label{fitparam}
\renewcommand{\arraystretch}{1}
\addtolength{\arraycolsep}{-1.0pt}
\end{table}
\begin{table}[h]
\renewcommand{\arraystretch}{1.5}
\addtolength{\arraycolsep}{3pt}
$$
\begin{array}{|c|c|c|c|}
\hline \hline
       \mbox{Parameters}    & c_1 (\mbox{GeV$^{-1}$}) & c_2 (\mbox{GeV$^2$})&c_3 (\mbox{GeV$^{-1}$})    \\
\hline
  \mbox{Values} &-0.006\pm0.002&-6.78\pm1.89&3.79\pm1.02 \\
                        \hline \hline
\end{array}
$$
\caption{Parameters appearing in the fit function of the coupling
form factor for $\Sigma_cND$ vertex.} \label{fitparam1}
\renewcommand{\arraystretch}{1}
\addtolength{\arraycolsep}{-1.0pt}
\end{table}
\begin{table}[h]
\renewcommand{\arraystretch}{1.5}
\addtolength{\arraycolsep}{3pt}
$$
\begin{array}{|c|c||c|c|}
\hline \hline
     \mbox{Coupling Constants}     & g_{\Sigma_bNB}(Q^2=-m_{B}^2) (\mbox{GeV$^{-1}$})&  g_{\Sigma_cND} (Q^2=-m_{D}^2) (\mbox{GeV$^{-1}$})  \\
\hline
  \mbox{Values} &12.96\pm3.49&3.78\pm1.06 \\
                         \hline \hline
\end{array}
$$
\caption{Values of the coupling constants $g_{\Sigma_bNB}$  and $g_{\Sigma_cND}$.}
\label{couplingconstant}
\renewcommand{\arraystretch}{1}
\addtolength{\arraycolsep}{-1.0pt}
\end{table}

To sum up, in this work, the strong  coupling constants  among the
heavy bottom spin--1/2 $\Sigma_b$ baryon, nucleon and $B$ meson as
well as the heavy charmed spin--1/2  $\Sigma_c$ baryon, nucleon and
$D$ meson, namely $g_{\Sigma_bNB}$ and $g_{\Sigma_cND}$, have been
calculated in the framework of the three-point  QCD sum rules.  The
obtained results can be applied in the analysis of the related
experimental results at LHC. The predictions can also be used in the
bottom and charmed mesons clouds description of the nucleon that may
be applied for the explanation of the exotic events observed by
different experiments. These results may also serve the purpose of
analyzing of the results of  heavy ion collision experiments like
$\overline{P}ANDA$ at FAIR. The obtained results may also come in handy
in the determinations of the changes in the masses, decay constants
and other parameters of the $B$ and $D$ mesons in nuclear medium.

\section{Acknowledgment}
This work has been supported in part by the Scientific and Technological
Research Council of Turkey (TUBITAK) under the research project 114F018.

\end{document}